\documentclass[runningheads]{llncs}
\usepackage[T1]{fontenc}
\usepackage{graphicx}
\usepackage{tabularx}
\usepackage[utf8]{inputenc}
\usepackage{array}
\begin{document}
\title{Overcoming Challenges in Agile and DevOps Integration: A Qualitative Study}
\titlerunning{Overcoming Challenges in Agile and DevOps Integration}
%
\author{Juliana Fraislebem  \inst{1} \and
Mali Senapathi\inst{2}\orcidID{0000-0003-3083-8069} \and
Michael Neumann\inst{3}\orcidID{0000-0002-4220-9641} \and
Eva-Maria Schön\inst{1}\orcidID{0000-0002-0410-9308}}
\authorrunning{J. Fraislebem et al.}
%
\institute{University of Applied Sciences Emden/Leer, Faculty of Business Studies, Emden, Germany
\email{eva-maria.schoen@hs-emden-leer.de}\\
\and
Auckland University of Technology, Auckland, New Zealand\\
\email{mali.senapathi@aut.ac.nz}
\and
University of Applied Sciences \& Arts Hannover, Hannover, Germany 
\email{michael.neumann@hs-hannover.de}\\
}
%
\maketitle              
\begin{abstract}
In response to the growing reliance on Agile and DevOps methodologies for enhancing software delivery speed and quality, this study investigates the persistent challenges and viable solutions associated with their integration. Although Agile promotes iterative development and customer responsiveness, and DevOps emphasizes automation and operational efficiency, their convergence in practice often presents significant organizational, structural, and technical hurdles. This research employs a qualitative methodology grounded in semi-structured interviews with six seasoned industry professionals across Brazil and Germany, each with extensive experience in both Agile and DevOps domains. The study identifies four core categories of integration challenges: Cultural
\& Organizational Barriers, Structural Constraints, Process \& Method Complexity, and Technical Limitations. Additionally, it offers four major solution domains: Team Structure \& Autonomy, Culture \& Collaboration, Process \& Change Management, and Automation \& Infrastructure. The findings underscore the importance of cultural alignment, proactive monitoring, automation, and
other practices in mitigating integration friction. The results contribute to a deeper understanding of the Agile-DevOps interface and offer practical insights for software organizations seeking to navigate this complex transition effectively.

\end{abstract}
\section{Introduction}
In today’s fast-paced and highly digitized business environment, software delivery has become a critical enabler of innovation and competitiveness. Organizations across sectors are increasingly turning to Agile and DevOps methods to respond more rapidly to customer needs and reduce time to market~\cite{Gall03092022}. Agile software development emphasizes iterative progress, customer collaboration, and responsiveness to change~\cite{Panchal_Shah_Shah_Bhatt_Tiwari_Yadav_2024}, while DevOps promotes collaboration between development and operations teams through automation, continuous integration, and continuous delivery~\cite{AMARO2025107583}. The integration of these two paradigms promises enhanced efficiency, faster deployment cycles, and improved product quality~\cite{WIEDEMANN2023100474}.

Agile is a software development methodology rooted in values such as customer collaboration, adaptive planning, and incremental delivery. It includes frameworks like Scrum~\cite{Schwaber2020} and Kanban~\cite{Anderson2010} that aim to create flexible and transparent workflows~\cite{SALTZ2020}. DevOps, by contrast, is not merely a set of tools or practices, but a cultural and organizational transformation. It combines software development and IT operations to shorten system development while providing features, fixes, and updates frequently and reliably~\cite{SMEDS2015}. DevOps emphasizes automation, continuous feedback, monitoring, and shared responsibilities across traditionally siloed teams~\cite{Gall03092022} and can be used for critical infrastructure with some adaptations (e.g., DevSecOps~\cite{Ramaj_2023}).
Despite the theoretical synergies between Agile and DevOps, organizations face significant challenges when attempting to integrate these methodologies into practice~\cite{AMARO2025107583}. Misalignment in team structures, cultural resistance, inconsistent feedback mechanisms, and fragmented toolchains hinder effective integration~\cite{Panchal_Shah_Shah_Bhatt_Tiwari_Yadav_2024}, ~\cite{NAYANAJITH2024}. Moreover, Agile focuses on flexibility and iterative planning, while DevOps emphasizes operational stability and automation, a combination that can create tension rather than harmony~\cite{AOUNI2025107569}. While many organizations are eager to reap the benefits of integration, they often lack a coherent framework and a clear understanding of the challenges involved~\cite{SMEDS2015}. Maturity and adoption models have emerged, but their evaluation remains uneven and their operationalization inconsistent~\cite{JAYAKODY2023}. To better understand and address these integration difficulties, this study investigates the following research questions (RQ): 
\begin{itemize}
    \item \textbf{RQ1:} What challenges do practitioners face when integrating Agile methodologies with DevOps practices?
    \item \textbf{RQ2:} What strategies and tools have proven effective in addressing these Agile-DevOps integration challenges in professional software development settings?
\end{itemize}
This paper is structured as follows: Section~\ref{sec2:RelWork} reviews the existing literature on Agile-DevOps integration, synthesizing known challenges and proposed strategies. Section~\ref{sec3:ResearchDesign} outlines the qualitative research methodology adopted in this study. Section~\ref{sec4:Results} presents the empirical findings, followed by a discussion of the results and the limitations in Section~\ref{sec5:DiscussionLimit}. Finally, Section~\ref{sec6:ConclFutWork} concludes the paper with key insights and directions for future research.

\section{Related Work}
\label{sec2:RelWork}
This section is based on peer-reviewed academic publications that examine the integration of DevOps with Agile methodologies. This section highlights the most relevant contributions that analyzed both Agile and DevOps.

A study of DevOps integration in Agile organizations is offered by Wiedemann et al.~\cite{WIEDEMANN2023100474}, using control theory to identify tensions between development and operations. These tensions stem from differing priorities, as Agile emphasizes speed and flexibility, while operations prioritize stability and risk mitigation. To address them, the authors propose a control model based on shared product vision, cross-functional ownership, and collaborative leadership to align priorities and reduce silos.


Panchal et al.~\cite{Panchal_Shah_Shah_Bhatt_Tiwari_Yadav_2024} identify key obstacles, including automation complexity, cultural misalignment, organizational inertia, fragmented communication, and tensions between Agile flexibility and DevOps continuous delivery. They also highlight performance evaluation challenges due to the lack of hybrid metrics and added complexity from security integration (DevSecOps). As a solution, the authors propose the Multivocal Software Factor Prioritization Development Approach (MSFPDA), prioritizing speed, quality, collaboration, and automation.


El Aouni et al.~\cite{AOUNI2025107569} examine the integration of Agile, DevOps, and Cloud paradigms. Despite this broader scope, they identify challenges relevant to Agile-DevOps integration, including cultural misalignment, fragmented workflows, and toolchain complexity. A key issue is the lack of unified frameworks, leading to inconsistent coordination and fragmented testing practices. To address these, the authors advocate for cultural integration aligned with the CALMS model (Culture, Automation, Lean, Measurement, Sharing), supported by shared tools and cross-functional Agile frameworks such as Scrum. They emphasize that mindset and cultural cohesion are critical for successful integration.

Amaro, Pereira, and Silva~\cite{AMARO2025107583} identify key challenges in Agile-DevOps integration, including limited coverage of early lifecycle phases, lack of standardized practices, and inconsistent use of operational metrics and feedback. These issues result in misaligned planning and automation, fragmented toolchains, and conflicting evaluation criteria. To address this, they propose a capability mapping framework aligning 37 DevOps capabilities with 30 IEEE lifecycle phases. They also recommend cross-functional roles, unified KPIs, and integrated toolchains to improve traceability, collaboration, and lifecycle alignment.

In a practical analysis, Samarawickrama and Perera~\cite{SAMARAWICKRAMA2017} identify gaps between Scrum and DevOps. While Scrum supports planning and management, it lacks mechanisms for continuous integration, deployment, and operational feedback. DevOps, in contrast, emphasizes automation but lacks a comprehensive process framework. These gaps result in misalignment and delayed production feedback. To address this, the authors propose Continuous Scrum, integrating CI/CD, automated testing and deployment, centralized version control, and monitoring into the Scrum process.

Senapathi, Buchan, and Osman~\cite{Senapathi.2018} present a case study of DevOps adoption in an Agile organization, highlighting challenges such as skill shortages, limited training, and resistance to change. The scale of technical and cultural change, alongside technological hurdles (e.g., cloud migration, CI/CD, monitoring), often led to burnout and slowed adoption. Ambiguities in responsibility also caused coordination gaps. To address these issues, the organization integrated operations staff into Scrum teams, adopted CI/CD and collaboration tools, and decentralized infrastructure ownership to promote autonomy, accountability, and cultural alignment.

Finally, Agarwal, Gupta, and Choudhury~\cite{AGARWAL2018} highlight technical challenges in Agile-DevOps integration, particularly toolchain fragmentation that disrupts continuity and feedback loops essential to Agile workflows. Manual processes in code review, deployment, and rollback slow delivery and reduce responsiveness. The absence of real-time feedback and insufficient automation in testing and security further hinder rapid iteration and early validation. To address these issues, the authors recommend unified DevOps pipelines, continuous monitoring, and integrated platforms to enhance feedback, collaboration, and alignment.

In summary, prior research identifies recurring challenges in Agile-DevOps integration, including cultural tensions between agility and operational stability~\cite{WIEDEMANN2023100474}, organizational and process misalignment~\cite{AOUNI2025107569,Gall03092022}, and technical barriers such as toolchain fragmentation and proliferation~\cite{AGARWAL2018,Panchal_Shah_Shah_Bhatt_Tiwari_Yadav_2024}. Proposed solutions include control models for managing tensions~\cite{WIEDEMANN2023100474}, capability mapping frameworks aligning DevOps with lifecycle processes~\cite{AMARO2025107583}, and process improvement models for adoption~\cite{SALTZ2020}. Nevertheless, the literature remains fragmented and provides limited empirical evidence on implementing these frameworks in practice~\cite{Senapathi.2018,SMEDS2015}. This gap motivates our study.

\section{Research Design}
\label{sec3:ResearchDesign}
To answer our research questions this study employed a qualitative research design based on semi-structured interviews~\cite{ResearchProtocol.2025}. The methodology follows established practices in qualitative social research and applies the framework of qualitative content analysis according to Mayring~\cite{MAYRING2000,MAYRING2022}. 

\subsection{Data Collection}
\textit{Development of the Interview Guide:} To enable structured but open-ended conversations, an interview guide was developed~\cite{ResearchProtocol.2025}. First, a broad list of relevant questions and topics was compiled based on the research goals. These were then critically examined, ensuring that questions were open-ended, understandable, and non-leading. The remaining items were arranged in a logical sequence to establish a clear interview structure. Topics were grouped into overarching thematic blocks to support coherence and transparency. A test interview was conducted with a pilot participant not included in the study sample to validate the clarity and relevance of the guide.

\textit{Execution of the Interviews:} The six interviews were conducted between 21.05.2025 and 16.06.2025, via Zoom conferencing, in a one-on-one format between the interviewer and each participant. Each interview lasted between 30 and 50 minutes. The participants responded in English or Portuguese~(\ref{tab:ParticipantsDemog}). With informed consent, all sessions were recorded and subsequently transcribed verbatim. All participant quotes originally in Portuguese were translated into English by the author. To preserve semantic fidelity, only quotations were translated by the first author (a native Portuguese speaker fluent in English). All analyses used the original Portuguese text to maintain meaning and tone.

\newcolumntype{Y}{>{\raggedright\arraybackslash}X}
\newcolumntype{C}[1]{>{\centering\arraybackslash}p{#1}}
\begin{table}[t]
\centering
\footnotesize
\renewcommand{\arraystretch}{1.15}
\setlength{\tabcolsep}{4pt}

\resizebox{\textwidth}{!}{%
\begin{tabular}{|c|l|l|l|c|c|c|}
\hline
\textbf{P} & \textbf{Country / Language} & \textbf{Company Type} & \textbf{Role} & \textbf{Total} & \textbf{Agile} & \textbf{DevOps} \\
\hline
P1 & Germany / English & Freelancer & Agile Coach & 24 & 17 & 15 \\
\hline
P2 & Brazil / Portuguese & Tech Company & Engineering Director & 24 & 20 & 20 \\
\hline
P3 & Germany / English & IT Consulting Company & Consultant / Product Owner & 20 & 13 & 10 \\
\hline
P4 & Brazil / Portuguese & State-owned Public Company & Developer / Architect & 20 & 11 & 11 \\
\hline
P5 & Germany / English & IT Consulting Company & Managing Director (Dev/Arc/DevOps) & 30 & 15 & 9 \\
\hline
P6 & Germany / English & IT Consulting Company & Consulting Director / Team Lead & 10 & 9 & 5 \\
\hline
\end{tabular}%
}
\caption{Participant demographics (experience in years)}
\label{tab:ParticipantsDemog}
\end{table}

\subsection{Sample and Participants}
The sample consisted of six experienced industry professionals selected through purposive sampling, ensuring a diverse and comprehensive representation of Agile-DevOps integration experience. The inclusion criteria were: (1) a minimum of five years of total experience in software development or related roles; (2) practical experience with both Agile methodologies and DevOps practices; (3) direct involvement in Agile-DevOps integration efforts in real-world software projects.

The participants brought diverse perspectives from a range of organizational settings, industries, and professional roles. All had over 10 years of experience, including 9 to 20 years in Agile and 5 to 20 years in DevOps, reflecting both strategic and hands-on involvement. This expertise provided a solid basis for exploring Agile-DevOps integration. Of the six participants, two were based in Brazil and four in Germany, offering cross-cultural insights. An overview of their roles and experience is provided in Table~\ref{tab:ParticipantsDemog}.

\subsection{Data Analysis}
The interview transcripts were analyzed using qualitative content analysis following Mayring~\cite{MAYRING2000,MAYRING2022} and in alignment with our research protocol~\cite{ResearchProtocol.2025}. We combined a deductive phase, guided by insights from the Agile–DevOps literature, with an inductive phase capturing themes that emerged directly from participants’ accounts. MAXQDA supported data management and coding, and its integrated AI functions were used to assist the analysis. Coding units were defined as complete responses to individual questions. We then consolidated the deductive and inductive results into a hierarchical code–subcode structure. The final structure is visualized in the hierarchical code-subcode models presented later in Figure 1 (Challenges) and Figure 2 (Solutions). Finally, we synthesized findings across interviews to identify commonalities, recurrent patterns, and outliers, yielding a structured view of the Agile–DevOps integration landscape.

\section{Results}
\label{sec4:Results}
This section presents the findings from the qualitative analysis of the interview data. Results are organized into two subsections: Identified Challenges and Proposed Solutions.

\subsection{Identified Challenges to Agile-DevOps Integration}
\begin{figure*}[htbp]
\includegraphics[scale=0.54]{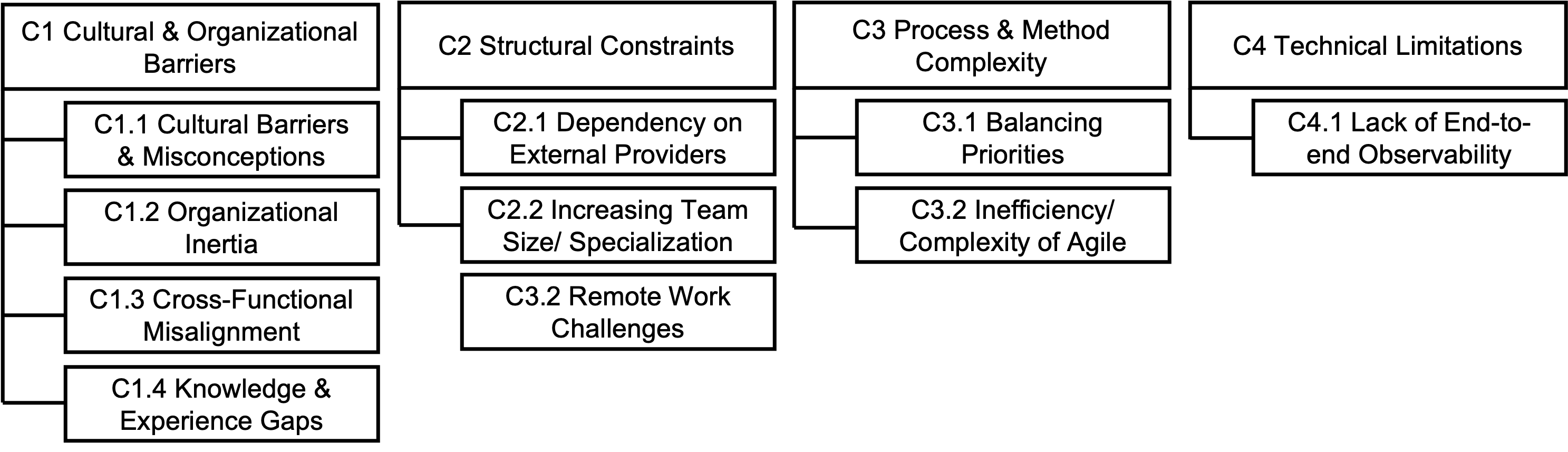}
\caption{Overview of Challenges to Agile-DevOps Integration}
\label{fig:CodeHierarchyChallenges}
\end{figure*}

The challenges encountered in integrating Agile and DevOps practices in the case setting were grouped into four overarching categories through thematic coding: (C1) Cultural \& Organizational Barriers, (C2) Structural Constraints, (C3) Process \& Method Complexity, and (C4) Technical Limitations. These categories and their associated subthemes are illustrated in Figure~\ref{fig:CodeHierarchyChallenges}.

\textbf{\textit{C1 Cultural \& Organizational Barriers.}} Cultural and organizational barriers emerged as the most frequently cited category of challenges, with 27 coded references across all six interviews. These barriers represent a significant obstacle to aligning Agile and DevOps practices with existing organizational norms, values, and ways of working.

\textit{C1.1 Cultural Barriers \& Misconceptions:} Many participants described misconceptions about Agile and DevOps. Agile was often seen as a rigid process rather than an adaptable mindset, while DevOps was reduced to tooling instead of collaboration and feedback. These oversimplifications led to mismatched expectations, especially when teams rigidly applied Scrum patterns in DevOps contexts.

“\textit{It’s mostly a combination of agile seen as a rigid structure to follow. Funny enough, that’s exactly the opposite of what Agile is about. […] In my experience, whenever it’s not a perfect fit [Agile \& DevOps], it is usually because people do not really understand what either is}”. – P1

A related misconception was the belief that implementing tools such as GitLab or CI/CD pipelines alone constituted DevOps adoption. Participants noted that effective DevOps requires customization and dedicated personnel, not just off-the-shelf solutions. 


Ingrained habits and adherence to familiar processes reinforced resistance to change, particularly among experienced stakeholders. These cultural dynamics often slowed the shift toward more agile, cross-functional, and iterative ways of working. Participants also pointed to an underlying “blame culture” that further hindered the shift toward agility and experimentation.

\textit{C1.2 Organizational Inertia:} Participants highlighted how organizational structures and slow decision-making processes hindered transformation. While agile principles could be implemented at the team level, systemic alignment with HR, finance, or procurement was often lacking. This "glass wall" limited the impact of retrospectives and impeded continuous improvement. Even when issues were repeatedly identified, teams lacked the authority to address them.


\textit{C1.3 Cross-Functional Misalignment:} Participants highlighted persistent separation between teams, both in terms of mindset and organizational structure, as a major obstacle to effective collaboration. Development, operations, infrastructure, and business teams often worked in silos, with limited shared ownership or coordination. This disconnect led to inefficiencies, miscommunication, and unresolved production issues.


\textit{C1.4 Knowledge \& Experience Gaps:} Participants described gaps in both technical knowledge and hands-on experience as key barriers to adopting Agile and DevOps. Developers from traditional environments often lacked familiarity with operational practices such as monitoring and troubleshooting, making the shift to shared responsibility particularly challenging. Conversely, infrastructure specialists without coding backgrounds struggled to engage with development concerns. These gaps were amplified when onboarding new hires unfamiliar with internal tools and workflows.

“\textit{You have a developer who came from an environment where they weren’t even allowed access to production. They’re used to this model of: I’ll just code here, throw it over the wall, and someone else will deploy it at some point. Typically, this developer doesn’t have much knowledge about infrastructure. So, if overnight you flip the switch and say: ‘Now you’re responsible for this’, it’s like, wow, what are the best practices for setting up proper monitoring? How do I use these tools? [...] There’s a skill gap in troubleshooting.}” – P2

\textbf{\textit{C2 Structural Constraints.}} Structural constraints were cited as practical and organizational factors that limited the effective implementation of Agile and DevOps principles. Though less frequently mentioned than cultural barriers, these challenges high-lighted the friction caused by dependencies and growing internal complexity.

\textit{C2.1 Dependency on External Providers:} Participants described how reliance on external actors, such as app marketplaces or third-party hosting providers, limited their teams’ agility. For instance, deploying updates through platforms like Google or Apple introduced delays, approval gates, and limited rollback options, forcing teams into more risk-averse practices. In other cases, production environments were fully managed by external vendors, leaving teams without access or control. This prevented them from making changes, troubleshooting, or experimenting without explicit customer approval, undermining DevOps principles of shared responsibility and operational autonomy.

“\textit{When you roll out a new version of an app, you don’t have the autonomy to push it to all your users. So anytime you want to make a change, no matter how small, it has to go through an approval process. […] If you make a mistake that’s actively hurting your business, you’re forced to wait. That delay might be hours, or it might be days. So you realize that this strategy of 'fail fast and recover quickly' just doesn’t work here. You have to avoid the error, because the cost of failure is enormous.}” – P2

\textit{C2.2 Increasing Team Size \& Specialization:} One participant noted that the growing complexity of modern software development requires larger, more specialized teams. While specialization improves technical depth, it also increases coordination over-head and complicates communication. Teams must now align across roles like cloud, testing, security, and frameworks, making it harder to maintain shared understanding, visibility, and cohesive delivery without strong coordination mechanisms.


\textit{C2.3 Remote Work Challenges:} A participant noted that remote work introduced structural friction that impacted collaboration and visibility between teams. Without in-person interactions, it became harder to understand roles, align on shared goals, or resolve issues, especially in newly formed or cross-functional teams.


\textbf{\textit{C3 Process \& Method Complexity.}} Participants described challenges stemming from delivery methods themselves, particularly when workflows or frameworks were rigid, overly complex, or poorly adapted to the team’s context. These issues led to confusion and friction, especially when teams were expected to follow formalized agile rituals without flexibility.

\textit{C3.1 Balancing Priorities:} A key concern raised by participants was the difficulty of balancing competing demands in a DevOps environment. Teams struggled to divide time between operational support, such as bug fixing, handling urgent requests, and assisting surrounding teams, and the pursuit of longer-term development goals such as innovation and feature delivery.


\textit{C3.2 Inefficiency \& Complexity of Agile Methods:} Some participants expressed concern about how agile frameworks were implemented. While acknowledging that approaches like Scrum or SAFe provide useful concepts, they were often applied too rigidly or scaled in ways that added unnecessary complexity. Particularly in large corporate environments, participants described a tendency to adopt standardized frameworks such as SAFe, which often added overhead and undermined agility.


\textbf{\textit{C4 Technical Limitations.}} Participants also highlighted system-level and infrastructural issues that limited the effectiveness of DevOps implementation. These technical barriers directly impacted core DevOps principles such as automation, observability, and rapid feedback. 

\textit{C4.1 Lack of End-to-End Observability:} In service-oriented architectures, a participant described how the absence of unified observability made troubleshooting significantly more difficult. When errors occurred, identifying the root cause required navigating multiple systems and manually correlating logs. This fragmentation delayed resolution and increased cognitive load during incident response. A more integrated view of system behavior, combining real-time monitoring and traceability across services, was considered essential for fast recovery, proactive maintenance, and effective incident management.


\subsection{Identified Solutions to Agile-DevOps Integration}
\begin{figure*}[htbp]
\includegraphics[scale=0.53]{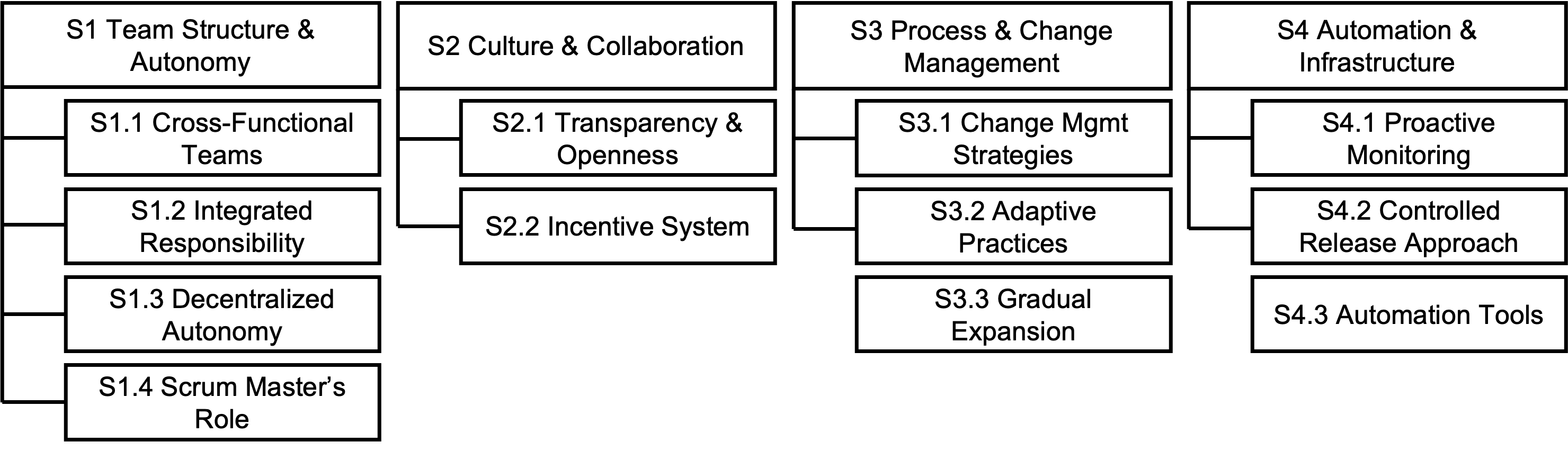}
\caption{Overview of Solutions to Agile-DevOps Integration}
\label{fig:CodeHierarchySolutions}
\end{figure*}

To address the integration challenges between Agile and DevOps, participants pro-posed a range of practical and cultural solutions, which are organized into four main categories (see Figure~\ref{fig:CodeHierarchySolutions}). These categories reflect both structural and behavioral enablers that support alignment across teams, processes, and technologies.

Notably, S1 Team Structure \& Autonomy and S4 Automation \& Infrastructure were the most frequently discussed solution areas, with 15 and 14 coded mentions respectively. S3 Process \& Change Management also featured prominently, indicating a strong focus on adapting delivery models over time. The following sections present each solution theme and its associated practices in more detail.

\textbf{\textit{S1 Team Structure \& Autonomy.}} This category includes solutions related to how teams are structured and empowered within Agile-DevOps environments. The participants emphasized the importance of redistributing responsibilities, promoting autonomy, and bridging the traditional role divides. These practices were seen as enablers of faster feedback loops, ownership, and responsiveness.

\textit{S1.1 Cross-Functional Teams:} Cross-functional teams were described as essential for diagnosing and resolving issues in complex platforms spanning infrastructure, security, and application layers. By combining expertise, teams were able to identify root causes and implement practices such as load testing. This integration improved iterative problem-solving and stakeholder communication. However, building such teams required time and addressing knowledge gaps between roles. To mitigate this, some participants recommended starting with a pilot team and scaling gradually.


\textit{S1.2 Integrated Responsibility:} Participants emphasized that DevOps success relies on shared responsibility across the software lifecycle. The “you build it, you run it” mindset fostered accountability and improved quality, as developers became more attentive to operational implications when they became responsible for maintaining what they built. 

“\textit{In my experience, when developers are also responsible for operations, they develop differently, because if something breaks, they’re the ones who might get woken up at 3 a.m. to fix it. No one wants that. So, they test more carefully, they pay more attention to performance. If you just hand it off, and it explodes in production, the ops team is left dealing with something they didn’t build. That’s my main advice: integrate operations into development not as a function, but as a responsibility. Let developers deal with what happens in production, it will absolutely lead to better code and higher quality overall.}” – P3

\textit{S1.3 Decentralized Autonomy:} Autonomy at the team or individual level was seen as a key enabler of agility, particularly in continuous delivery environments. Developers needed the ability to deploy, monitor, and roll back changes independently, without approval chains, to react quickly to issues and reduce delivery delays. Infrastructure tooling enabled this independence, while shared leadership and pull-based workflows reinforced individual ownership.

“\textit{I need the tooling so that if a developer deploys something and notices an issue, they can quickly revert that change. I can’t have them depending on approvals or needing to talk to another team. The person who owns the service, who built it and deployed the change, must have the autonomy to roll it back immediately.}” – P2

\textit{S1.4 Scrum Master's Role:} The Scrum Master was highlighted as crucial in facilitating communication and resolving cross-team issues. By coordinating with external stakeholders, removing blockers, and fostering collaboration, this role supported alignment, particularly in distributed teams.


\textbf{\textit{ S2 Culture \& Collaboration.}} This category includes solutions aimed at improving the cultural and interpersonal foundations of Agile-DevOps. These enablers, though less tangible, were described as critical for fostering trust, breaking down silos, and enabling collaboration. Participants emphasized psychological safety, transparency, and team motivation as key factors.

\textit{S2.1 Transparency and Openness:} Participants highlighted transparency as a cornerstone of Agile-DevOps collaboration. Making work visible through tools such as JIRA or Planner helped align expectations, identify blockers, and create shared understanding. Retrospectives supported open discussion of issues, both within teams and with stakeholders. Beyond tools, success depended on a culture where problems could be raised openly and feedback was a regular practice.


\textit{S2.2 Incentive System:} A participant noted that incentive structures could either support or undermine Agile-DevOps goals. When rewards focused on individual performance or strict KPIs, collaboration and risk-taking often suffered. Tying incentives too closely to OKRs encouraged goal negotiation over meaningful outcomes. Instead, incentives should reinforce team contribution, shared accountability, and collaborative behavior.


\textbf{\textit{S3 Process \& Change Management.}} This category includes strategies for adapting processes, scaling Agile-DevOps practices, and managing change in complex environments. Participants mentioned tailoring practices to context, managing organizational complexity, and introducing change iteratively.

\textit{S3.1 Change Management Strategies:} Participants emphasized that implementing Agile and DevOps requires dedicated change management, especially in large organizations. Top-down support was seen as critical, with leadership setting the strategic direction and teams providing feedback from the ground. Bottom up efforts alone often struggled to gain traction across hierarchical structures. Participants also stressed the importance of honest self-assessment, noting that companies often overestimate their agility maturity. Effective change required stakeholder communication, training, and realistic expectations.


\textit{S3.2 Adaptive Practices:} Rather than rigidly applying frameworks like Scrum or SAFe, participants emphasized the importance of understanding the principles behind them. No single method fits all contexts, and factors such as deadlines, client technical maturity, and team experience required adaptation. Participants warned against rule-following without purpose and instead encouraged reflective practice, flexibility, and a broader understanding of change.

“\textit{Most people focus on the rules, they learn something new and think, ‘These are the rules, I have to follow them.’ But in Agile, what really matters are the principles behind those rules. For example, in Kanban, limiting WIP is just a tool, the real goal is to improve flow by removing blockers. The same goes for Scrum: the sprint isn’t meant to constrain you, but to help focus your effort in a positive way. So, the key is to ask: What am I trying to achieve? What principle does this rule support? And if a rule doesn’t seem to help, consider whether you’re missing the bigger picture or whether it’s time to adapt the framework to your context.}” – P1

\textit{S3.3 Incremental Deployment:} Incremental deployment was recommended as a low-risk strategy for delivering changes. By rolling out updates gradually to a subset of users, teams could identify issues early, reduce impact, and adjust before full release. This approach was seen as especially useful in high-stakes environments.


\textit{S3.4 Gradual Expansion:} Rather than attempting large-scale transformation, one participant advocated for starting with a well-defined team or product area. Demonstrating success locally, with frequent deployment, stable systems, and a collaborative culture, created a model that could be expanded across the organization. This approach lowered the risk of failure and created internal credibility for new ways of working. 


\textbf{\textit{S4 Automation \& Infrastructure.}} This category includes technical solutions that support Agile and DevOps practices at scale. Participants highlighted tooling, release management, and observability as key enablers of safe, repeatable, and high-quality delivery. While culture and structure were seen as foundational, many participants emphasized that without the right infrastructure, agility and speed remain aspirational.

\textit{S4.1 Proactive Monitoring:} Proactive monitoring and observability were described as essential for detecting performance issues early and accelerating resolution. Tools integrated with platforms like Grafana helped teams to trace API latency and identify system bottlenecks, a clear improvement over manual troubleshooting. 


\textit{S4.2 Controlled Release Approach:} To balance speed with reliability, participants described controlled release processes based on fixed cadences (e.g., weekly or biweekly), with defined cutoff points. Changes were then tested, certified, and gradually rolled out, often to a subset of users. This ensured predictable and stable deployments while maintaining frequent delivery. This strategy was especially relevant in contexts where external platforms, such as app stores, limited deployment control, as discussed in C2.1. In those cases, where rollbacks were difficult and approvals delayed, a fixed release cadence helped teams maintain predictability and reduce risk.

“\textit{We have a release process built around what we call a ‘train.’ If you want your change to go live, it needs to get on the train by a certain cutoff date. Everything in that batch goes through extensive testing, certification, and gradual rollout. If you miss the date, you wait for the next train, usually in a week or two. It’s not like continuous integration where you deploy constantly. With apps, you just don’t have that level of control, so we had to build a cadence that reduces risk.}” – P2

\textit{S4.3 Automation Tools:} Automation was described as the foundation for DevOps success. Participants emphasized the value of investing in high-quality tooling, particularly for test automation, CI/CD pipelines, and observability, to prevent long-term technical debt and enable faster, more reliable delivery cycles. Automated testing was particularly critical, although teams often struggled with test coverage, maintenance, and diagnosing failures. Some participants expressed optimism that AI and large language models (LLMs) may help reduce the effort required to build and maintain robust test suites.

“\textit{Another important challenge is building the habit of writing automated tests. Even now, in 2025, we shouldn’t still need to talk about why it matters, but in practice it’s still hard. […] It only works if the whole team buys in and knows how to write tests, and that’s much harder. […] Quality and speed both depend on automated testing. At my last job, we clearly saw that over time, automated tests allowed us to deliver faster without sacrificing quality.}” – P4


\section{Discussion and Limitations}
\label{sec5:DiscussionLimit}
This qualitative study provides insight into the challenges and solutions associated with integrating Agile and DevOps in professional software development. The findings confirm and expand on prior literature by offering real-world perspectives on integration issues and concrete strategies.

Among the identified challenges (RQ1), cultural and organizational barriers (C1) emerged as the most prominent, appearing in all six interviews (27 coded references), far exceeding the mentions of other categories. These barriers spanned misconceptions about Agile and DevOps, organizational inertia, cross-functional misalignment, and gaps in technical knowledge. Participants described misunderstandings, such as equating DevOps with tooling alone or interpreting Agile as rigid methodology, which often led to mismatched expectations and resistance to collaboration. These findings are in strong alignment with prior research, particularly by Wiedemann et al.~\cite{WIEDEMANN2023100474}, Senapathi et al.~\cite{Senapathi.2018}, and El Aouni et al.~\cite{AOUNI2025107569}, who emphasize the need for cultural alignment, shared understanding, and continuous learning in integration contexts.

Other challenges were less mentioned (4 mentions each for C2 and C3; 1 for C4). Structural constraints (C2), such as external dependencies and coordination overhead, limited team autonomy and responsiveness. These challenges are consistent with findings by Senapathi et al.~\cite{Senapathi.2018}, who highlight the friction caused by limited cross-team autonomy and centralized decision-making. Similarly, El Aouni et al.~\cite{AOUNI2025107569} emphasize that disconnected workflows and weak organizational alignment impair effective Agile-DevOps integration. 

Process and method complexity (C3) highlighted issues like the rigidity of scaled Agile frameworks and the difficulty of balancing operational demands with long-term development goals. This resonates with Samarawickrama and Perera~\cite{SAMARAWICKRAMA2017}, who argue that Scrum alone lacks the built-in mechanisms needed for operational integration, and that rigid adherence to frameworks can create additional friction. Lastly, the relative infrequency of technical limitations (C4), such as lack of observability or tooling gaps, suggests that these are often seen as symptoms of deeper organizational and process-level misalignments. This interpretation is supported by Agarwal et al.~\cite{AGARWAL2018}, who point out that fragmented toolchains and insufficient automation usually reflect unresolved structural or cultural issues within the organization.

To address these challenges, participants highlighted several strategies (RQ2). Key among them is the use of cross-functional teams with shared responsibility for development and operations. This approach was seen as essential for breaking down silos and improving end-to-end delivery accountability. It directly supports recommendations by Panchal et al.~\cite{Panchal_Shah_Shah_Bhatt_Tiwari_Yadav_2024} and Wiedemann et al.~\cite{WIEDEMANN2023100474} on role convergence and collaborative leadership. Participants also stressed the value of unified CI/CD pipelines, often using tools such as Jenkins, GitLab, and Docker, to ensure process continuity and transparency. These tools enable real-time feedback and reduce cycle times, which is essential for harmonizing Agile’s rapid delivery goals with DevOps’ emphasis on reliability. The relevance of such tools is well-supported in the literature by El Aouni et al.~\cite{AOUNI2025107569}, Acevedo et al.~\cite{Acevedo_2026}, and Agarwal et al. \cite{AGARWAL2018}, who highlight the need for integrated automation and shared monitoring platforms. 

Further mentioned practices included embedding operations staff in Scrum activities, promoting transparency and collaboration through shared dashboards, and reviewing operational metrics jointly. These practices resonate with the CALMS framework (Culture, Automation, Lean, Measurement, Sharing)~\cite{AOUNI2025107569} and are often cited in integration-focused research. Finally, participants emphasized that successful integration depends on leadership support, ongoing training, and a culture that fosters experimentation and tolerance for failure. This view echoes Wiedemann et al.~\cite{WIEDEMANN2023100474} and Senapathi et al.~\cite{Senapathi.2018}, who highlight leadership’s role in enabling learning, autonomy, and cultural transformation.

Although this study was conducted rigorously applying established guidelines, we have to point out limitations. First, the small sample size (six participants) limits the generalizability of the findings. While the sample was diverse in terms of roles and geographies, it may not fully capture the breadth of challenges faced across different industries or organizational maturities. Second, the qualitative nature of the analysis introduces the risk of interpretive bias, despite efforts to maintain rigor via structured coding. Lastly, while triangulation with existing literature was conducted, the study did not directly evaluate the effectiveness of proposed integration strategies in practice.

\section{Conclusion \& Future Work}
\label{sec6:ConclFutWork}

This study explored the integration of Agile methodologies and DevOps practices, focusing on practical challenges (RQ1) and the strategies used to address them (RQ2). Based on qualitative interviews and a review of the literature, we identified organizational, cultural, technical, and process-related barriers to integration.

The findings show that cultural resistance, role ambiguity, and fragmented toolchains remain key obstacles. Moreover, the misalignment between Agile’s iterative planning and DevOps’ emphasis on stability and automation further contributes to friction in practice.  However, the study also highlights concrete strategies that have proven effective in addressing these issues. These include the formation of cross-functional teams, shared responsibility for operations, investment in automation and monitoring tools, and leadership-driven change management. These results reinforce that successful integration requires not only technical solutions but also organizational and cultural transformation.

While limited by its small sample size and qualitative scope, this study provides valuable empirical evidence. Future research should extend this work through larger cross-industry and longitudinal studies assessing the long-term impact of integration strategies. Furthermore, the development of standardized frameworks and maturity models could further support Agile-DevOps transformation. Drawing on insights from industry experts, this research underscores the importance of context-sensitive, collaborative, and incremental approaches to Agile-DevOps integration, providing both academic and practical guidance.
%
%
%
 \bibliographystyle{splncs04}
 \bibliography{references}

\end{document}